\documentclass[preprint]{aastex61}

\usepackage{amsmath}
\usepackage{mathtools}
\usepackage{graphicx}
\usepackage[utf8]{inputenc}

\shorttitle{Absolute Magnitude Calibration for W UMa-Type Systems}
\shortauthors{Mateo \& Rucinski}

\begin{document}

\title{Absolute Magnitude Calibration for W UMa-Type Systems Based 
on \textit{Gaia} Data\footnote{This work has made use of data 
from the European Space Agency (ESA)
mission {\it Gaia} (\url{https://www.cosmos.esa.int/gaia}), processed by
the {\it Gaia} Data Processing and Analysis Consortium (DPAC,
\url{https://www.cosmos.esa.int/web/gaia/dpac/consortium}). Funding
for the DPAC has been provided by national institutions, in particular
the institutions participating in the {\it Gaia} Multilateral Agreement.} 
}

\author{Nicole M.\ Mateo}
\affiliation{Department of Astronomy and Astrophysics, 
University of Toronto\\50~St.~George St., Toronto, Ontario, M5S~3H4, Canada}
%\author[I-2724-2017]{Slavek M.\ Rucinski}    % invalid, why? this is as provided
\author{Slavek M. Rucinski} 
\affiliation{Department of Astronomy and Astrophysics, 
University of Toronto\\50~St.~George St., Toronto, Ontario, M5S~3H4, Canada}

\correspondingauthor{Slavek M. Rucinski}
\email{rucinski@astro.utoronto.ca}

\begin{abstract}
Tycho-Gaia Astrometric Solution (TGAS) parallax data are used to determine absolute magnitudes $M_V$ for 318 W~UMa-type (EW) contact binary stars. A very steep (slope $\simeq -9$), single-parameter ($\log{P}$), linear calibration can be used to predict $M_V$ to about 0.1 -- 0.3 mag over the whole range of accessible orbital period, $0.22\!<\!P\!<\!0.88$ days. A similar calibration for the most common systems with $0.275\!<\!P\!<\!0.575$ days predicts $M_V$ values to about 0.06 -- 0.16 mag. For orbital period values both shorter and longer than the central range, the period dependence is respectively steeper and shallower, i.e.\ the binaries are fainter in $M_V$ than predicted by the whole-range, linear law. The steepness of the relation for short-period systems implies important consequences for the detectability of the faintest binaries defining the short-period cut-off of the period distribution. Although the scatter around the linear $\log{P}$-fit is fairly large (0.2 -- 0.4 mag), the current data do not support the inclusion of a $B\!-\!V$ color term in the calibration. 
\end{abstract}


\section{Introduction} 
\label{sec:intro}

W~UMa-type (EW) binaries are short-period ($P < 1$ day) binaries consisting of eclipsing, solar-type stars. They are characterized by their continuous brightness variability without well defined moments of eclipse egress and ingress; the eclipses have equal depths, indicating that both components have similar effective temperatures. Currently, the most accepted model of their structure utilizes the Roche geometry of two orbiting stars sharing a common convective envelope which can be described by a common equipotential \citep{Lucy1968a,Lucy1968b}. This makes them externally simple objects since the complexities of the internal mass and energy exchange mechanisms are hidden within the envelope. With the observed prevalence of unequal component masses, EW binaries can be envisaged as single structures where the more massive star provides most of the luminosity and the less massive star, because of its large orbit relative to the mass center, carries most of the angular momentum. On color-magnitude diagrams, they appear as objects close to the Main Sequence (MS), though one or both components may have highly evolved internal structures \citep{Step2006a,Step2006b,Step2009,Step2011}\footnote{The astroph/0701529 version of the paper \citet{Step2006b} should be consulted for the correct expression for the adopted angular-momentum loss law (St\c{e}pie\'{n}, priv.\ information).}. 

The period--color correlation discovered by \citet{Egg1967} was one of the first indications that EW binaries may be similar to MS objects in terms of the effective temperature and the size scaling. Their apparent simplicity was the motivation for the absolute magnitude calibration of \citet{Rci1994}, though parallax and photometric data were relatively poor at the time. For full generality, the calibration used a two-parameter dependence on the orbital period ($P$) and the color ($B\!-\!V$ or $V\!-\!I$) to represent the binaries' mass progression along the MS as well as the considerable scatter away from it. The color index, however, came from photometric data of moderate quality and the period--color correlation was not explicitly taken into account.

The parallax information provided by the {\it Hipparcos\/} Mission \citep{esa1997} led to an improved calibration by \citet{RD1997}. It utilized 40 EW binaries with a period range of $0.24\!<\!P\!<\!1.15$ d, a color range of $0.26\!<\!B\!-\!V\!<\!1.14$, and an absolute magnitude range of $1.4\!<\!M_V\!<\!6.1$. The data were still of modest quality with the need for two parameters remaining unresolved.
The calibration was later broadened to globular-cluster EW binaries \citep{Rci2000}, with a notable lack of an obvious metallicity ($[Fe/H]$) dependence (we note that metallicity is irrelevant in the present context of nearby, disk population EW binaries; see \citet{Rci2013}). The calibration proved to be a useful tool in verifying memberships of EW binaries in open   \citep{Mazur1995,MK1999,MK2000,Maciej2008,Chen2016}
and globular \citep{KTK1997,JKa1997,ITh1999} clusters. Notwithstanding, the \citet{RD1997} calibration still suffered not only from the limited precision of the parallax data, but also from the poorly standardized photometric data as well as the underestimation of the frequency of occurrences of additional stellar components \citep{Trip3}.

This study attempts to improve the absolute magnitude calibration using the first {\it Gaia\/} satellite Data Release (DR1) \citep{Gaia2016a,Gaia2016b}. It also includes many systems discovered recently by systematic sky surveys. The DR1 has been based on astrometric solutions called TGAS (Tycho-Gaia Astrometric Solutions) which combine positional information from the {\it Hipparcos\/} and the {\it Tycho-2\/} catalogs with the new {\it Gaia\/} data to isolate the proper motion components, thus improving the parallax determinations. 

The new dataset for EW binaries is superior to the previous {\it Hipparcos\/}-based one and provides a much better calibration which we describe below. In addition to the determination of new parameters entering the $M_V$--period--color calibration, our goal was to verify the need for the two independent parameters, $\log P$ and $B\!-\!V$, which are known to be correlated via the \citet{Egg1967} relation. As independent variables, $\log P$ and $B\!-\!V$ are very different in terms of observational accuracy -- the orbital period is normally known with a relative accuracy typically $10^{-5}$ higher than the photometric data. Additionally, the photometric data suffer not only from simple measurement and standardization errors (which, when combined, are currently at the 0.01--0.03 mag level) but also from interstellar reddening corrections (which are comparably uncertain for typical distances of a few hundred parsecs). 

The uniformity and the completeness of the photometric data have improved substantially during the intervening years since the \citet{RD1997} paper. The {\it Hipparcos\/} catalog listed many previously undetected, low-amplitude EW binaries down to $V\!<\!7.5$ \citep{Rci2002} which have then been studied in detail in the following years. EW detections, which used to be accidental and random, became more systematic thanks to the {\it ASAS\/} survey \citep{asas} providing standardized $V$ magnitudes (here used as $V_{\rm max}$, i.e.\ at light-curve maximum) for EW binaries within the range $8\!<\!V\!<\!13$ for a large fraction of the sky. In addition, a systematic photometric survey by the {\it Tycho-2\/} project \citep{Hog2000} provided uniformly calibrated $B\!-\!V$ color indices for stars down to $V\simeq10$. 
Temporal variability monitoring and planetary transit surveys such as Super-WASP \citep{Norton2011,Lohr2013}, OGLE  \citep{Sosz2015}, and CRTS \citep{Marsh2017}\footnote{For a list of W~UMa-binary searches based on these data, see \citet{Marsh2017}.} have obtained data of high photometric accuracy, but they are typically over smaller sections of the sky than {\it ASAS\/} and at fainter magnitudes than the range relevant for this study (the \textit{Gaia} targets are all brighter than $V \simeq 12$, see below). It should be noted that the discoveries of low brightness systems are particularly important in determining the spatial density of the intrinsically faintest systems close to the sharp period cut-off at about 0.2 d (see \citet{Rci2007} and references therein). 

The current paper is an attempt to combine the new parallax data from the {\it Gaia\/} satellite with results from photometric surveys available at this time. Section~\ref{data} presents the parallax, photometric, and reddening data while Section~\ref{res} discusses the new calibrations. Finally, Section~\ref{concl} summarizes the results of this paper.

\section{Observational data used}
\label{data}

\subsection{Parallaxes}

The motivation for this study is the availability of newly measured, high precision parallaxes from the \textit{Gaia} mission as well as standardized $UBV$ photometry for W~UMa binaries; the details of the photometric database are described in the next sub-section. Cross-matching of the catalogs led to the selection of EW binaries from \citet{RD1997}, \citet{asas}, \citet{GGM2006}, and \citet{Terr2012}, provided that they have corresponding \textit{Gaia} parallaxes. After removing the poorer data points, the final sample consists of 318 EW binaries with absolute-magnitude errors $\sigma (M_V) \!<\! 0.5$ mag\footnote{The systems from \citet{RD1997} that are not included in this study due to their errors $\sigma (M_V)\!>\!0.5$ mag: TY~Men, HI~Pup, RW~Dor, MW~Pav, RS~Col.}. We used the standard formula for absolute magnitudes: $M_V \!=\! 5\log(\pi) + V_{max} - 3.1 \times E_{(B\!-\!V)} - 10$ with $\sigma (M_V)$ estimated through propagation of errors contributed by the parallax and the photometry. 

The parallax determination via TGAS is a substantial improvement to that of the {\it Hipparcos\/} sample (HIP) \citep{RD1997}. The median parallax error for the HIP sample was 1.30 mas (milli-arcsec) whereas it is 0.28 mas for the TGAS sample. Similarly, the median {\it relative\/} error $\sigma (\pi)/\pi$ has been reduced from 0.13 to 0.038. Since the magnitude scale $M_V$ is logarithmic, the latter value is of greater importance. From the HIP sample of 40 systems, 30 have data in the TGAS database\footnote{The HIP binaries absent from the TGAS database: S~Ant, V752~Cen, V759~Cen, VW~Cep, RW~Com, $\varepsilon$~CrA, XY~Leo, UZ~Oct, V566~Oph, GR~Vir.}. This rather high fraction of HIP systems absent from the TGAS database may indicate frequent occurrences of perturbing companions to these binaries \citep{Trip3} which would have introduced difficulties in the astrometric solutions.  

\begin{figure}
\begin{center}
\includegraphics[width=12.0cm]{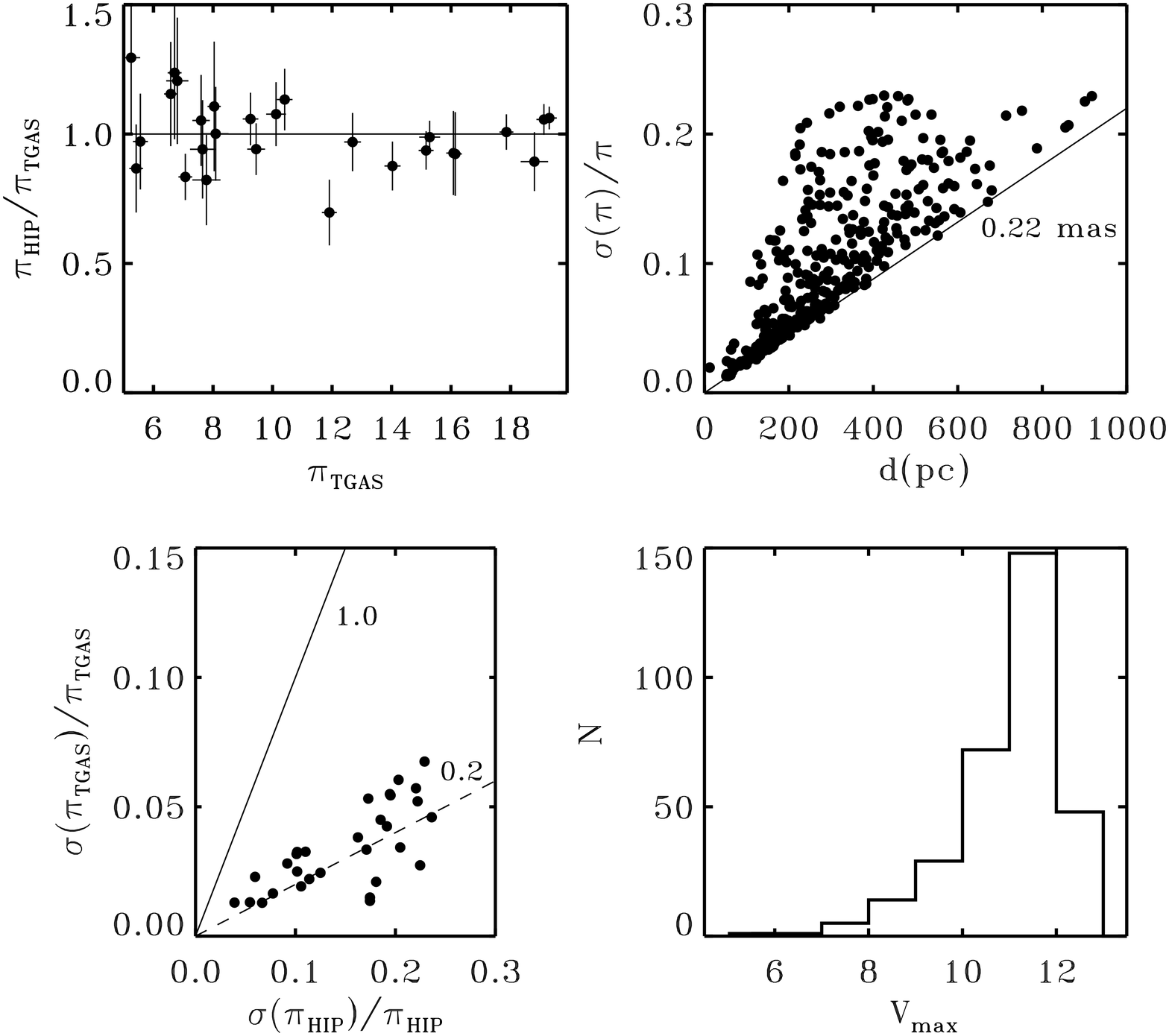}  
\end{center}
\caption{
\footnotesize 
The general properties of the TGAS sample of 318 W~UMa-type binaries. 
{\it Upper left:\/} Comparison of the parallax determinations for 30 binaries 
in common with the {\it Hipparcos \/} sample. 
The ratios $\pi_{\rm HIP}/\pi_{\rm TGAS}$ versus $\pi_{\rm TGAS}$ are shown. 
{\it Lower left:\/} The relative errors $\sigma(\pi)/\pi$ for the HIP 
and the TGAS samples. The TGAS data appear to be about 5 times more 
accurate than the HIP data. 
{\it Upper right:\/} The relative errors $\sigma(\pi)/\pi$ versus the 
distance $d \!=\! 1000/\pi_{\rm TGAS}$, with $\pi_{\rm TGAS}$ in mas
(milli-arcseconds). Note that the data imply a typical minimum TGAS
error of $\sigma(\pi) \simeq 0.22$ mas (the sloping line).
{\it Lower right:\/} The number of binaries per maximum brightness 
$V_{\rm max}$ in 1-magnitude intervals. The number of W~UMa binaries in 
the TGAS sample drops sharply for $V_{\rm max}\!>\!12$, a limitation due to
the limited depth of the {\it Tycho\/} data. 
}
\label{fig_gen_prop}
\end{figure}

\placefigure{fig_gen_prop}   % Fig.1 general properties of the sample

The HIP and TGAS results for common systems appear to be in mutual agreement as shown in Figure~\ref{fig_gen_prop}; the graph presents the properties of the input parallaxes for both datasets. The TGAS relative errors $\sigma (\pi)/\pi$ are approximately 5 times smaller than those of HIP which results in a five-fold increase in accessible distances $d \!=\! 1000/\pi$ (with $\pi$ in mas or milli-arcsec units) and thus a larger volume by about 125 times. The apparent scaling of the minimum relative errors with the distance (see Figure~\ref{fig_gen_prop}) implies a typical minimum parallax error of about 0.22 mas for the TGAS sample. 

As pointed by \citet{LK1973}, relative errors of $\sigma (\pi)/\pi \!>\! 0.05$ may result in biased determinations of distances. As seen in Figure~\ref{fig_gen_prop}, this may have happened to a large fraction of the TGAS sample at distances greater than about 200 pc. Thus, while the new sample includes more distant binaries than the HIP sample, the quality of the parallax data in terms of $\sigma (\pi)/\pi$ for distant objects in both samples is comparable and moderate. However, thanks to its much larger sample volume, the TGAS dataset includes more stars within a wider range of parameters. We made sure that the binaries observed with lower accuracy contributed less to the final $M_V$ calibration solutions using a weighting scheme with weights $\propto 1/{\sigma^2}_{(M_V)}$.

The nearest system 44~Boo (i~Boo, HIP~73695) does not have a \textit{Gaia} parallax but its {\it Hipparcos\/} measurement has been included in the current sample because of the importance of its small relative parallax error: $\pi_{HIP}\!=\!79.95 \pm 1.56$ mas. The value $(B\!-\!V)\!=\!0.94\pm0.02$ has been taken from \citet{FM2000} while the value $V_{max}\!=\!5.87\pm0.02$ has been taken from \citet{Hill1989}. This object is the only one in our sample with a HIP number in place of a TYC number.

\subsection{Photometric data}

In selecting the photometric data, $V_{\rm max}$ and $B-V$, we decided not to use scattered photometric information published in numerous, single-object papers. Recently, discoveries have become more organized due to large-scale, systematic photometric surveys. We intentionally chose precision over potentially improved accuracy by preferentially using these survey data. 

The sky has been uniformly surveyed for stellar variability with photometric errors $\!<\!0.01$ mag (enabling discoveries of variability with amplitudes of about 0.05 mag) by the {\it Hipparcos\/} satellite down to $V \!<\! 7.5$ \citep{Rci2002}. {\it Tycho-2\/} magnitude and
color data \citep{esa1997} are reliable to about $V \simeq 10$ (see Fig.~2 in \citet{Rag2010}), but the survey may be incomplete in terms of variability detections. Calibrated photometry for large parts of the sky reaching down to $V\!<\!15$ has been undertaken by a few wide-sky surveys. In addition to the {\it Hipparcos\/} and the {\it Tycho-2\/} catalogs, data for $V_{\rm max}$ have been taken from three other sources that we judged to be large enough to provide sufficient uniformity: \citet{GGM2006},
%(hereinafter called GGM2006), 
\citet{asas}\footnote{\url{http://www.astrouw.edu.pl/asas/?page=catalogs}}, and \citet{Terr2012}. We re-evaluated the errors for $V_{\rm max}$ and assumed that they are not smaller than 0.01 mag due to residual calibration errors. 

Uniformly calibrated $B\!-\!V$ data for EW binaries have been taken primarily from \citet{Terr2012} which analyzed 606 binaries with $V_{\rm max} \!<\! 14$.
%\footnote{In addition, since \citet{Terr2012} 
%is a more recent photometric study, we quoted the paper's values 
%for $V_{max}$ and $(B\!-\!V)$ for ten HIP systems: 
%AB~And, TZ~Boo, AC~Boo, CK~Boo, SX~Crv, AP~Leo, UV~Lyn, V839~Oph, VZ~Psc, V781~Tau.}. 
No directly measured colors exist for about a third of the binaries with measured parallaxes and readily available $V_{\rm max}$ values. For such cases, we used the infrared indices $J\!-\!H$ and $H\!-\!K$ from the 2MASS survey \citep{2mass} to estimate the $B\!-\!V$ colors. While this may be a risky conversion since the color indices for EW binaries do not necessarily obey the MS relations, this approach assures a high degree of uniformity within our data. Using the extensive tabulation of \citet{PM2013}\footnote{\url{http://www.pas.rochester.edu/~emamajek/EEM\_dwarf\_UBVIJHK\_colors\_Teff.txt}}, the agreement between the independent color transformations from $J\!-\!H$ and $H\!-\!K$ to $B\!-\!V$ gave estimates for the uncertainties of the derived $B\!-\!V$ values. 

\subsection{Interstellar reddening}

The interstellar extinction ($A_V$) and reddening ($E_{\rm B\!-\!V}$) corrections must be included because their neglect would create systematic biases to $V_{\rm max}$ and $B\!-\!V$. For EW binaries, these corrections are expected to be relatively small for distances of tens to a hundred parsecs, but become larger and very uncertain beyond a few hundred parsecs due to the very non-uniform distribution of interstellar matter in the local Galactic Disk.

For uniform treatment of the interstellar corrections, we utilized the distances derived from the new \textit{Gaia} parallaxes and assumed an exponential density decay with galactic height to approximate the interstellar matter distribution. For this, we closely followed \citet{Nat2013} but modified the paper's approach (developed for fields at the center of the Galaxy) in the following way: instead of assuming that interstellar extinction ``hits the wall'' at the Galactic Bulge, we assumed that the maximum values are determined by integrating to infinity as in \citet{Schleg1998}. In general, for any distance $R$ and galactic latitude $b$:
\begin{equation}
A_V\!=\!\int_0^R \exp(-r\,|\!\sin(b)|/H)\,\mathrm{d}r,
\end{equation}
with the scale height $H\!=\!164$ pc \citep{Nat2013}. Here, the upper limit can be the actual distance to the binary ($R\!=\!d$) or it can be infinity ($R \!=\! \infty$), with the latter giving $A_V^{max}$ as in \citet{Schleg1998}. From the ratio of the definite integrals to both values of $R$:
\begin{equation}
A_V \!=\! A_V^{max} \times (1-\exp(-d\,|\!\sin(b)|/H)).
\end{equation}
Finally, we assumed that the relation between the reddening and absorption corrections is the standard law, $A_V\!=\!3.1 E_{\rm B\!-\!V}$.

All data used in this paper are listed in Table~\ref{tab_data}. The table is arranged by increasing orbital period. 

\placetable{tab_data}       % Table 1 with the input data

\subsection{Limitations of the sample} 
\label{limit}

The current sample has some limitations that impact the present attempts to obtain a luminosity calibration: (1)~the decreased precision of the parallax data for the most distant objects (Figure~\ref{fig_gen_prop});  (2)~the loss of about one fourth of the binaries due to an apparently more stringent criteria (than those of the HIP sample) when the {\it Tycho\/} and the {\it Gaia\/} data were combined into TGAS; and (3)~our neglect of the potential presence of companions to the EW binaries. We comment on them in turn, bypassing \#1 since it is a rather obvious one. 

\begin{figure}
\begin{center}
\includegraphics[width=12.0cm]{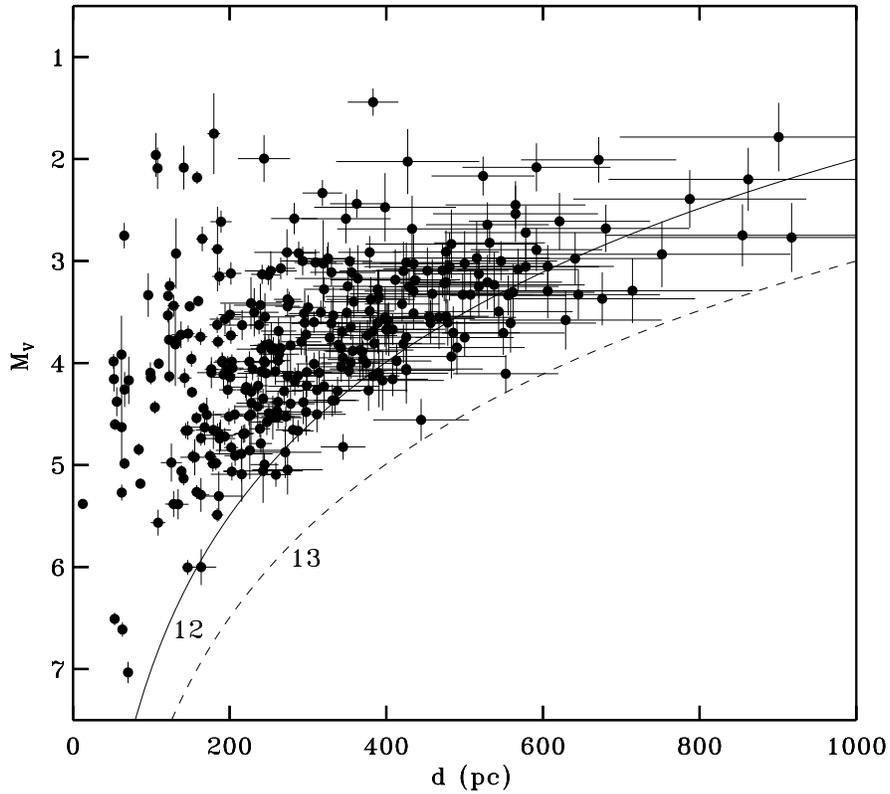}  
\end{center}
\caption{
\footnotesize 
The $M_V$ range accessible to the calibration is severely limited for 
intrinsically faint EW binaries. The plot shows the $M_V$ determinations versus 
the distance $d$ in parsecs. The limits imposed by the depth 
are shown for $V_{\rm max}\!=\!12$ (continuous curve) and  
$V_{\rm max}\!=\!13$ (broken curve). 
Note the increased distance errors for large distances due to progressively 
larger values of the relative errors $\sigma(\pi)/\pi$.
}
\label{fig_dist_Mv}
\end{figure}

Due to limitation \#2, the current sample cannot be used to estimate the spatial density of EW binaries. For the faintest systems, the volume currently explored is particularly small. These are the binaries with the shortest orbital periods which, similar to previous studies, remain under-represented in spatial counts. This is illustrated in Figure~\ref{fig_dist_Mv} in a plot of accessible luminosities (expressed in $M_V$) versus the distance. The limits for the apparent magnitudes $V\!=\!12$ and $V\!=\!13$ are shown; they roughly correspond to the depth of the {\it Tycho\/} data which defined the TGAS data selection via the initial-epoch positions for the proper-motion eliminations. The decrease in the accessible range of $M_V$ magnitudes may combine with (or possibly modify) the apparent sharp drop in spatial density of EW binaries taking place at about $M_V \!>\! 5$ \citep{Rci2006,Rci2007}. As shown in Figure~\ref{fig_gen_prop}, the healthy increase in binary numbers versus $V_{\rm max}$ stops beyond $V \simeq 12$. Thus, the definition of the short-period cut-off for EW binaries remains an open question due to the lack of data for the faintest systems.

\placefigure{fig_dist_Mv}          % depth and M_V. Fig.2

We are very much aware of the fact that many EW binaries have companions whose presence may affect the $M_V$ and the $B-V$ photometric data. Targeted searches \citep{Trip1,Trip2,Trip3} found that additional components are very common with a frequency approaching one hundred percent. Unfortunately, the searches had limited depths with only the brightest EW binaries (usually $V < 10$) having been successfully investigated. With this in mind, we decided not to correct for the brightness contributions of these companions since adjusting only the apparently brightest systems could potentially introduce a systematic luminosity bias in the sample; we simply assumed that the presence of individual binary companions would show up in the increased $\sigma(M_V)$ errors.

Finally, in evaluating the $V_{\rm max}$ values, we entirely neglected slow trends in mean luminosity \citep{RP2002} observed in about one third of the EW binaries \citep{Marsh2017}. These very recently discovered trends will no doubt be studied in the coming years, but at this moment we again had to assume that they will simply increase the $\sigma(M_V)$ errors.

\section{Results}
\label{res}

\subsection{Definitions and general considerations}

The previous absolute magnitude calibrations for EW binaries were developed as linear, three-parameter relations of the form $M_V \!=\! a_0 + a_P \log{P} + a_C (B\!-\!V)_0$. \citet{Rci1994} simplistically argued that the expected calibration may have such a shape, possibly with a contribution from the mass-ratio ($q$) which is usually unknown without spectroscopic studies. Although it was recognized that the period--color correlation discovered by \citet{Egg1967} possibly plays a role through the correlation of $\log{P}$ and $B-V$, it was not explicitly included in these considerations. The improved data from the HIP sample seemed to confirm the need to include both the period and the color, but the significance of a solution with two independent parameters was low though fully explainable by the moderate quality of the parallax data.

The two terms in the previous calibrations, $a_P$ and $a_C$, relate to the quantities which are not only mutually correlated but are also determined with very different uncertainties. Thanks to the repetitive nature of photometric observations, the orbital period is usually known with very high accuracy compared to the color which is usually relatively inaccurate due to a large fraction of EW binaries never having been observed in standard photometric systems. 
In conjunction, the individual calibrations to standard systems of many binaries give results with discrepancies at the level of a few hundredths of a magnitude, indicating the presence of systematic errors in the color calibrations. Two additional physical causes produce further uncertainties: (1)~EW binaries do change their colors with the orbital phase (for $B\!-\!V$, it is typically by 0.02--0.05 mag) -- a variation interpreted within the Lucy model as due to gravity distributions over the common equipotential; and (2)~interstellar extinction systematically reddens the colors but is poorly determined in individual cases. As a result, the colors are seldom known to better than a few hundredths of a magnitude which makes it a much weaker independent variable than the period. 

\placefigure{fig_Mv_logP}     % M_V vs. logP

\placetable{tab_fits}              % Table 2 with coefficients

% changed  paragraphs    =>

Our least-squares fits have been done in succession starting from (1)~the simplest linear fit utilizing $\log{P}$ as the independent variable, (2)~by splitting the full $\log{P}$ range into three linear sub-sections, (3)~by considering a quadratic dependence, and finally (4)~by supplementing the relation with the de-reddened color index $(B-V)_0$ as the second independent variable. Figure~\ref{fig_Mv_logP} shows the $M_V$ vs.\ $\log{P}$ dependence while the coefficients obtained for the fits are given in Table~\ref{tab_fits}. We used the standard least-squares regression to evaluate the values of $\chi^2$, the reduced $\chi_r^2$ (per degree of freedom), the mean weighted standard deviation $\epsilon$ and the mean weighted absolute deviation $\delta$, with weights for the individual stars $\propto 1/{\sigma^2}_{(M_V)}$. Because the intrinsic scatter in $M_V$ is unknown, we used bootstrap experiments to estimate the coefficient errors. The experiments consisted of 10,000 samples with repetitions from the population of $n$ stars (for the whole sample, $n\!=\!318$). The coefficient uncertainties were estimated as the 15.8th and the 84.1st percentiles of the resulting coefficient distributions relative to the median (50th percentile); such scatter levels correspond to the mean standard errors of a Gaussian distribution. To improve the quality of the least-squares fits, we shifted the independent variables $X$ and $Y$ to the approximate centres of their respective ranges, as given in Table~\ref{tab_fits}. We measured the quality of each fit by calculating the coefficient of determination, $R^2$, which is defined as the fraction of variance reproduced by the model, $R^2 = 1 - SS_{\rm res}/SS_{\rm tot}$, where $SS_{\rm res}$ is the sum of the residual (unexplained) variances computed by using local, non-parametric smoothing, while $SS_{\rm tot}$ is the sum of the total variances. Thus $0 \le R^2 \le 1$, where 1 indicates the best model fit and 0 indicates the fit not represented by the assumed model.  

\subsection{Individual solutions} 

\begin{figure}
\begin{center}
\includegraphics[width=12.0cm]{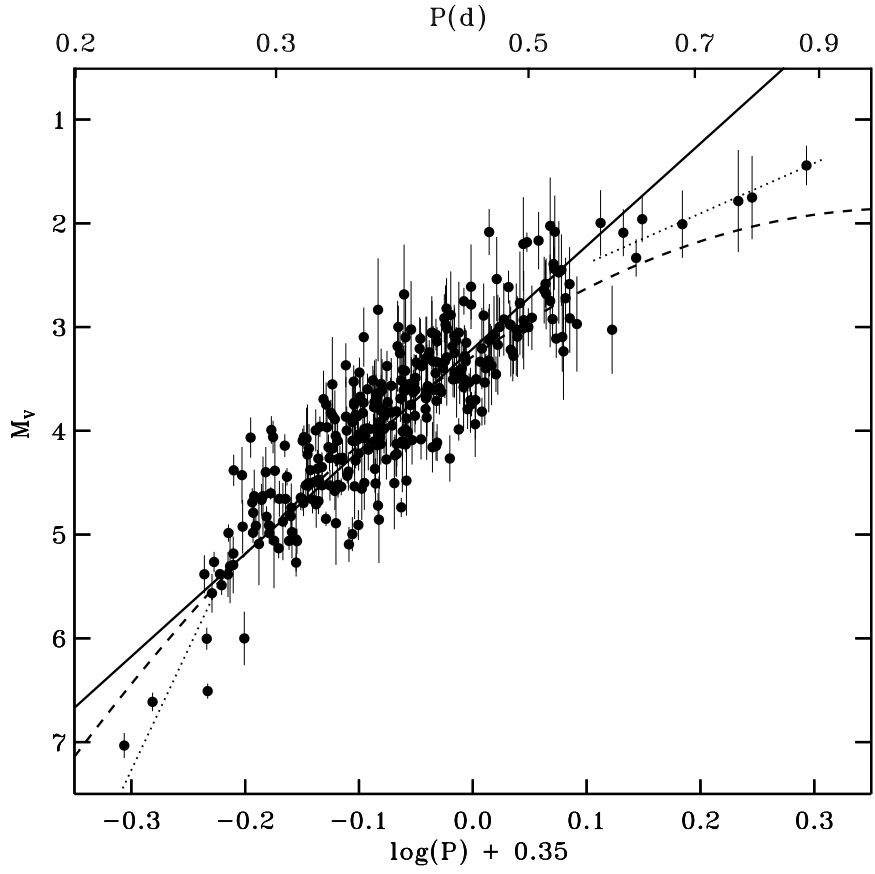}  
\end{center}
\caption{
\footnotesize 
The observed $M_V$ vs.\ $\log{P}$ relation together with several fits as
in Table~\ref{tab_fits}. The lines show the individual solutions: the linear 
fit for the whole range $0.22 \!<\! P \!<\! 0.88$ d (the straight continuous line), 
the quadratic fit (the curved broken line), and two separate linear fits 
for $P\!<\!0.275$ d and for $P\!>\!0.575$ d (the dotted lines). 
Note that departures from the general linear fit at both ends are 
highly significant, even exceeding $10 \sigma (M_V)$ in individual cases.    
The orbital period values (in days) corresponding to the  
abscissa $X = \log{P} + 0.35$ are given along the upper horizontal axis. 
}
\label{fig_Mv_logP}
\end{figure}

We discuss below the calibration fits in relation to Table~\ref{tab_fits} which gives the numerical details. Figure~\ref{fig_Mv_logP} illustrates the fits, as explained in its caption.
\begin{description}
\item[Solution \#1] The linear fit over the whole period range, $0.22\!<\!P\!<\!0.88$ d, is obviously an oversimplification. The fit shows systematic trends in the residuals (see Figure~\ref{fig_Mv_logP}) for the short period and the long period ends where EW binaries are fainter than what the linear regression predicts. The trends are reflected in the value of $R^2 = 0.71$. The very steep dependence at the short-period end and the shallow dependence at the long-period end were noted before \citep{Rci2006,Paw2016} and may at least partially be explained by the shift of the dominant flux away from the $V$-band filter transmission.
\item[Solutions \#2 and \#3] The separate linear fits for the short-period end, $0.275\!<\!P$ d, and the long-period end,  $P\!<\!0.575$ d, are based on very few stars. Thus, no systematic trends in the solutions are visible and $R^2$ could not be reliably determined. We note that the values of $\chi^2_r$ are very different for the two solutions: while $\chi^2_r = 1.0$ for the long-period segment (\#3), $\chi^2_r = 19.3$ for the short-period segment (\#2). The large value of $\chi^2_r$ may be explained either by our under-estimation of the $\sigma(M_V)$ errors, by an inappropriate linear model (oversimplified description of the functional dependence of $M_V$ on binary parameters), or by an intrinsic scatter in $M_V$ (e.g. due to the neglected companions or the slow luminosity trends, Sec.~\ref{limit}). We return to this subject in Section~\ref{concl}.
We note that the particularly steep relation for the short-period systems is driven mostly by the new, high-precision parallax determinations for CC~Com and V523~Cas; the steepness of the dependence may have important implications on the detectability of short-period systems.
\item[Solution \#4] This is the best defined among the single-parameter, $\log{P}$-dependent solutions. It covers the central range $0.275\!<\!P\!<\!0.575$ d and is based on a large number of stars. Its better definition, relative to \#1, is reflected in the decreased value of $\chi^2_r$ from 6.2 to 4.9 and the increased value of $R^2$ from 0.71 to 0.84. 
\item[Solution \#5] A quadratic fit for the whole period range is only slightly better than the linear fit \#1. It does not reproduce the ends very well and it only leads to a marginal improvement in $\chi^2$ and $R^2$.
The added quadratic term was found to be insignificant through the $F$-test: the change from $\chi_2^2\!=\!1962$ to $\chi_3^2\!=\!1807$ results in $F \!=\! 0.0003$.  
 \end{description}

\begin{figure}
\begin{center}
\includegraphics[width=12.0cm]{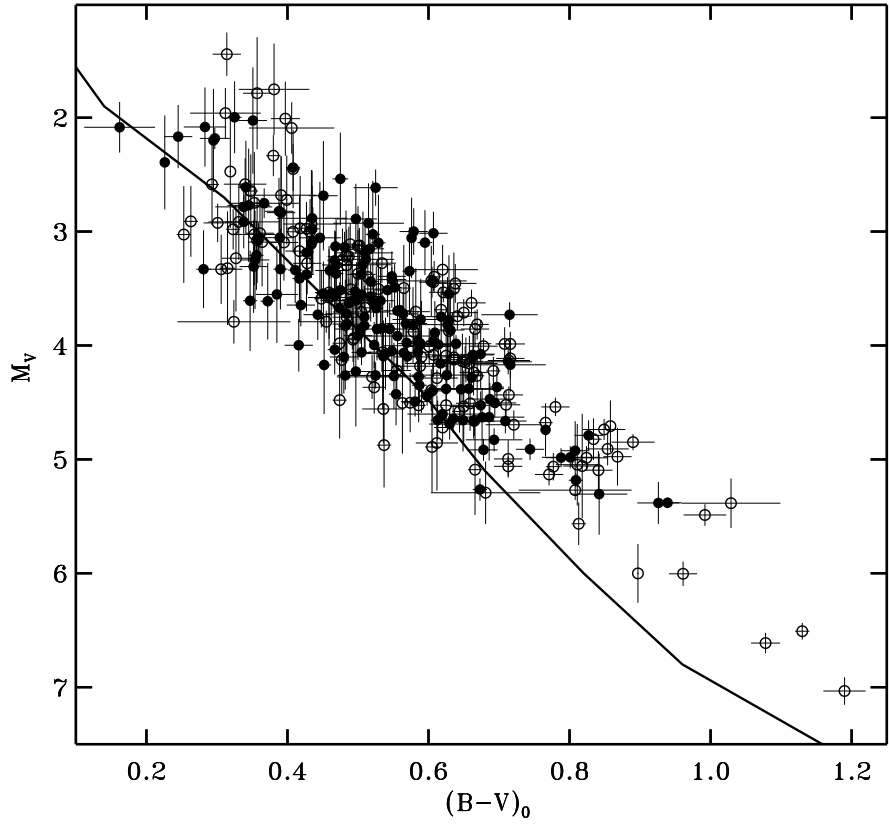}  
\end{center}
\caption{
\footnotesize 
The figure shows the color--magnitude diagram (CMD) for the current sample. 
The filled and open circles correspond to binaries which are either 
brighter (filled circles) or fainter (open circles) than the whole-range linear 
dependence (\#1 in Table~\ref{tab_fits}) shown in Figure~\ref{fig_Mv_logP}. 
There is no obvious separation between the two in the central grouping 
which contains the most typical EW binaries. At the ends, particularly the faint end, 
the open symbols dominate because of the curvature 
of the $M_V$ vs.\ $\log{P}$ dependence. 
The continuous line is the observational Main Sequence for solar
metallicity stars \citep{Gray2005}. 
}
\label{fig_Mv_BV}
\end{figure}

\placefigure{fig_Mv_BV}     % M_V vs. color: CMD

Before discussing the two solutions utilizing the de-reddened color index $(B-V)_0$,  \#6 and \#7, we present the familiar color--magnitude diagram (CMD) for the solar neighborhood EW binaries (Figure~\ref{fig_Mv_BV}). The EW binaries are located mostly above and to the right of the Zero Age Main Sequence. While evolved cores of one or both components may produce an increase in luminosity, the Lucy model predicts the main shift to occur in the color. With the observed tendency of unequal masses in EW binaries ($q \neq 1$), the secondary component should add almost no light but a lot of surface area which must result in a lower effective temperature. For details on the expected shifts in magnitude and color versus the mass ratio $q$, see Figure~8 in \citet{RD1997}. 
\begin{description}
\item[Solution \#6] This fit is basically solution \#1 with the added color term. It has the same form as the calibrations in \citet{Rci1994} and \citet{RD1997}. Since the color term has been added to improve the fit over the whole period range, we attempted to visualize a possible coupling of the $M_V$ deviations with the color by marking in Figure~\ref{fig_Mv_logP} the binaries that deviate from the linear period dependence \#1. An obvious correlation is visible at the short-period end (for $M_V\!>\!5.5$) where all the binaries are red and fainter than what the linear fit predicts. While fit \#6 is definitely better than fit \#1 and shows an expected reduction in $\chi^2$, the formal $F$-test gives $F\!=\!0.004$ and does not support the need for two variables; the deviating ends are too poorly populated to influence the solution. The slope coefficients are somewhat similar to those in \citet{RD1997}, but not identical which probably reflects the different proportion of the included short-period systems in the respective samples.
\item[Solution \#7] The fit for the central part of the period distribution ($0.275\!<\!P\!<\!0.575$ day) with the color term included is particularly informative. With the short-period and the long-period ends removed, the color term becomes entirely redundant. Thus, the inclusion of the short-period, red binaries which deviate in luminosity from the linear $\log{P}$ dependence resulted in the bias in the previous calibration \citep{RD1997} as well as in solution \#6.
 \end{description} 

 % <= changed paragraphs

In conclusion, we feel that solution \#4 is the best for representing the majority of detected EW binaries. For short-period and long-period binaries, the best solutions are currently \#2 and \#3, but future studies may result in a better functional form that could cover the whole period range. Finally, the color term is apparently not needed at all.

\section{Conclusions}
\label{concl}

The absolute magnitudes of 318 W~UMa-type binaries determined using their TGAS parallaxes from the \textit{Gaia} satellite show a rapid decrease in brightness as the orbital period becomes shorter -- an effect due to the combined change of the radiating area and of the effective temperature. A single, steep relationship $M_V \!=\! a_0 + a_1 \log{P}$, with $a_1 \simeq -9$, can serve as an approximate absolute-magnitude calibration for W~UMa-type binaries in the period range of 0.22 to 0.88 days and approximately over 5 magnitudes in $M_V$, predicting $M_V$ to about 0.08 mag around 0.45 d and to about 0.28 mag at both ends of the period distribution (solution \#1 in Table~\ref{tab_fits}; the prediction levels are based on the coefficient uncertainties). Since the potential presence of companions and of slow luminosity trends have been neglected, the precise determination of the fit coefficients is mostly due to the large size and the high consistency of the sample; in fact, individual deviations from the linear fit reach 0.2 -- 0.4 mag. 

When the period range is reduced to the most populated region within $0.275\!<\!P\!<\!0.575$ days, then the expected deviations from the linear $M_V=M_V(\log{P})$ predictions are reduced to 0.06 mag around 0.40~d and 0.16 mag at both ends of the region (solution \#4). This is a substantial improvement from the previous calibrations not only in terms of accuracy but also in terms of simplicity as a single-parameter dependence. Apparently, $\log{P}$ and the color $B-V$ correlate so well that the dependence can be reduced to a single-parameter one based solely on the very precisely determinable period; this may have  important implications for EW binary models (see below). The irrelevance of the color term is particularly important because of the generally inadequate color-index information for EW binaries. When checking the necessity for the color term, we attempted to circumvent the absence of a consistent color database by using only the targeted survey of \citet{Terr2012} and the transformed data from the 2MASS infrared survey \citep{2mass}. However, we recognize that the latter color transformations may have introduced their own systematic effects since the applicability of MS transformations to EW binaries was never verified. The previous $M_V$ calibrations that did require the color term were apparently biased due to the inclusion of faint, red, short-period binaries which appear to be fainter than a simple extension from the region 0.275 -- 0.575 d would predict. 

While binaries at the short-period ($P\!<\!0.275$ d) and the long-period ($P\!>\!0.575$ d) ends depart from the main relation and are systematically fainter, the current TGAS data are too sparse to provide definite determinations of the slopes for linear relationships. Nevertheless, we see an unquestionable steepening of the period dependence for the shortest-period binaries thus the available volume in the $V$-band for these stars shrinks considerably, emphasizing the necessity for red and near-infrared searches. The linear fits, in terms of $\chi^2$ values evaluated using observational $\sigma (M_V)$ errors, appear to worsen as one goes down the period sequence. Keeping in mind the small number of binaries at both ends, we note that we achieved the best linear fit for the systems with $P \!>\! 0.575$ d and the worst linear fit for the systems with $P\!<\!0.275$ d. This may be related to the recently documented slow trends in mean luminosity \citep{RP2002} observed in about one third of known EW binaries \citep{Marsh2017}. Such trends, which can reach about 0.1 mag, take several years to occur and may possibly be explained by magnetic phenomena. \citet{Marsh2017} found these tendencies to be less frequent among hotter binaries with radiative envelopes, which is in agreement with our better fit for the binaries with $P\!>\!0.575$ d.  

The simplicity of the period--luminosity relation may be considered as one of the confirming validity proofs of the  \citet{Lucy1968a,Lucy1968b} model. The model generated a vigorous discussion in the decade following its inception (mostly regarding its structural details) but not really reaching firm conclusions. Its geometrically and dynamically constrained nature was hard to reconcile with the demands for complex internal mass and energy exchange processes implied by its assumptions\footnote{For a highly readable historical account, see the review by one of its strong contributors, \citet{Webb2003}.}. However, it was recognized from the beginning that stable contact configurations can exist only when the components are sufficiently different in their internal structure (inhomologous) -- a condition which can be achieved most directly through evolution. Recent model analyses \citep{Step2006a,Step2006b,Step2009,Step2011} involve mass-exchange and mass-reversal evolutionary processes in stars that are at least moderately evolved. In view of the complexities of such processes, it is not excluded that the Lucy model is indeed the unifying factor providing one simple period--luminosity relationship despite a large range of initial parameters for pairs of stars expected to form EW binaries. Observationally, it is not obvious that the model is the only option for explaining EW binaries: the light curves carry a relatively modest information content while the only spectroscopically well observed EW binary at the moment, AW~UMa \citep{Rci2015}, appears to show a number of spectral discords with the Lucy model that point to the possibility of the binary being an Algol-type. With these reservations aside, our results indicate that in terms of the period--luminosity relation, the W~UMa binaries form a surprisingly uniform group of stars. Thus, either AW~UMa is an exceptional case of a semi-detached system mimicking a contact binary, or the spectral complexities observed at high spectral resolutions are indications of superficial, outer-atmospheric phenomena \citep{Eaton2016} which may be unrelated to these binaries' internal structure.  

\begin{acknowledgements}

The authors would like to express their gratitudes to Dr.\ Laurent Eyer for pointing out the imminent availability of the {\it Gaia\/} DR1 (TGAS) results and thus prompting them to undertake this study. As well, the authors would like to thank Dr.\ G. Pojma\'nski for his rapid replies to their inquiries regarding the ASAS project and Dr.\ K.\ St\c{e}pie\'{n} for commenting on the early version of the manuscript. 

This research has been supported by the Natural Sciences and Engineering Research Council of Canada to SMR. It has made extensive use of the SIMBAD and the VizieR databases (operated by the CDS, France) and of NASA's Astrophysics Data System (ADS). 

\end{acknowledgements}

% -------------------------------------------------------------------------------

% ============================================================
%                        Fig.1 properties of the sample          

\newpage

% Fig.1  ---------------------- properties ------------------------------

\newpage

% Fig.2 ---------------------- depth of the sample -----------------

% Fig.3 -------------------------- Mv - logP ---------------------

% Fig.4 -------------------------- CMD -----------------------------

% ===========================================================
% Tables

\newpage

% Table 1  data ----------------------------------------------------------

\begin{deluxetable}{lllrcccccrrrr}
%                   1234567890123

\tabletypesize{\footnotesize}    % 10 pts
\tablewidth{0pt}
\tablecaption{Observational data \label{tab_data}}
\tablenum{1}

\tablehead{\colhead{Name} & \colhead{TYC} & \colhead{$P$ (d)} 
         & \colhead{$V_{\rm max}$} & \colhead{$\sigma(V_{\rm max})$} & \colhead{$B\!-\!V$}
         & \colhead{$\sigma(B\!-\!V)$} & \colhead{$E_{\rm B\!-\!V}$} & \colhead{$\sigma(E(B\!-\!V))$}
         & \colhead{$\pi$ (mas)} & \colhead{$\sigma(\pi)$} & \colhead{$M_V$}
         & \colhead{$\sigma(M_V)$}
}
\startdata
    CC~Com &  1986-2106-1 & 0.220686 & 11.30 & 0.09 & 1.20 & 0.03 & 0.010 & 0.001 & 14.22 &  0.54 &  7.03 &  0.12 \\
  V523~Cas &   3257-167-1 & 0.233692 & 10.62 & 0.05 & 1.08 & 0.02 & 0.002 & 0.005 & 15.84 &  0.53 &  6.61 &  0.09 \\
    RV~Gru &   8446-662-1 & 0.259516 & 10.94 & 0.03 & 0.93 & 0.03 & 0.004 & 0.000 &  7.78 &  0.65 &  5.38 &  0.18 \\
    EI~CVn &   2548-936-1 & 0.260770 & 11.86 & 0.02 & 0.97 & 0.02 & 0.009 & 0.001 &  6.83 &  0.33 &  6.00 &  0.11 \\
    VZ~Psc &    581-259-1 & 0.261259 & 10.20 & 0.05 & 1.15 & 0.01 & 0.020 & 0.001 & 18.80 &  0.46 &  6.51 &  0.07 \\
  15867853 &   5507-705-1 & 0.263549 & 10.83 & 0.01 & 0.84 & 0.01 & 0.027 & 0.002 &  9.20 &  0.79 &  5.57 &  0.19 \\
    772054 &  4375-2406-1 & 0.264667 & 11.31 & 0.02 & 0.69 & 0.01 & 0.017 & 0.001 &  6.33 &  0.28 &  5.26 &  0.10 \\
    44~Boo &    HIP~73695 & 0.267818 &  5.87 & 0.02 & 0.94 & 0.02 & 0.001 & 0.002 & 79.95 &  1.56 &  5.38 &  0.05 \\
  V384~Ser &   2035-175-1 & 0.268739 & 11.87 & 0.02 & 1.01 & 0.03 & 0.018 & 0.004 &  5.43 &  0.23 &  5.49 &  0.09 \\
    MU~Aqr &  5178-1376-1 & 0.272218 & 11.12 & 0.02 & 1.06 & 0.07 & 0.031 & 0.002 &  7.45 &  0.74 &  5.38 &  0.22 \\
\enddata
\tablecomments{The star names, when fully numerical, correspond to the numbers 
in \citet{GGM2006}; in catalogs the numbers are referred to as {\it GGM2006-nnnn\/}. 
TYC are the numbers in the {\it Tycho\/} Catalog. All photometric data are in magnitude units,
while the parallax is in milli-arcsec (mas) = 0.001 arcsec.\\
The full version of the table is available electronically.
}
\end{deluxetable}

% Table 2  solutions -----------------------------------------------------

\begin{deluxetable}{lcccccccccc}

\tabletypesize{\footnotesize}    % 10 pts
\tablewidth{0pt}
\tablecaption{Calibration \label{tab_fits}}
\tablenum{2}

\tablehead{\colhead{Fit} & \colhead{Details} & \colhead{$a_0$} 
   & \colhead{$a_1$} & \colhead{$a_2$} & \colhead{n} & \colhead{$\chi^2$} 
   & \colhead{$\chi_r^2$} & \colhead{$\epsilon$} & \colhead{$\delta$} & \colhead{$R^2$}
}
\startdata
%\sidehead{\boldmath $M_V \!=\! a_0 + a_1 X$ }
%\cutinhead{\boldmath $M_V \!=\! a_0 + a_1 X$} 
    & { \boldmath $M_V \!=\! a_0 + a_1 X$ } &   &   &   &   &   &   \\[1.0ex]  % title
\#1 &
$0.22 \!<\! P \!<\! 0.90$ d&       &      &     &  318   &      &    \\  
 & $X\!=\! \log{P}+0.35$         
    & $3.21$                 & $-9.89$                     &     &     & 1962.2  & 6.2 & 0.40 & 0.31 & 0.71 \\ % LSQ
 & $P_0=0.4467$ d & $3.21_{-0.08}^{+0.08}$ & $-9.85_{-0.67}^{+0.61}$  &   &   &   &   \\[2ex]  % boot
\#2 &
$P \le 0.275$ d             &       &      &     &  12   &      &     \\  
    & $X\!=\! \log{P}+0.60$         
    & $6.11$                 & $-23.27$                     &     &     & 174.0  & 19.3 & 0.36 & 0.26 & (1.0) \\ % LSQ
 & $P_0=0.2512$ d & $6.15_{-0.19}^{+0.29}$ & $-22.37_{-7.73}^{+2.44}$   &   &   &   &   \\[2ex]  % boot
\#3 &
$P \ge 0.575 d$             &       &      &     &   9   &      &     \\  
    & $X\!=\! \log{P}+0.15$         
    & $1.90$                 & $-4.85$                     &     &     & 6.13  & 1.0 & 0.21 & 0.15 & (1.0) \\ % LSQ
 & $P_0=0.7079$ d & $1.91_{-0.09}^{+0.08}$ & $-4.54_{-0.86}^{+1.40}$   &   &   &  &   \\[2ex]  % boot
% central, my fit #6
\#4 &
$0.275 \!<\! P \!<\! 0.575$ d &       &      &     &  297   &      &    \\  
    & $X\!=\! \log{P}+0.40$         
    & $3.73$                 & $-8.67$                     &     &     & 1453.3  & 4.9 & 0.37 & 0.29 & 0.84 \\ % LSQ
 & $P_0=0.3981$ d & $3.73_{-0.06}^{+0.06}$ & $-8.67_{-0.64}^{+0.65}$   &   &   &   &  \\[2ex]  % boot
% quadratic, fit #4
    & { \boldmath $M_V \!=\! a_0 + a_1 X + a_2 X^2$ } &   &   &   &   &   &   \\[1.0ex]  % title
\#5 &
$0.22 \!<\! P \!<\! 0.90$ d &       &      &     &  318   &      &    \\  
    & $X\!=\! \log{P}+0.35$         
    & $3.28$                 & $-7.54$              &  +9.96  &     & 1806.8  & 5.8 & 0.38 & 0.30 & 0.82\\ % LSQ
& $P_0=0.4467$ d & $3.29_{-0.08}^{+0.08}$ & $-7.52_{-1.18}^{+1.31}$ & $+10.19_{-5.51}^{+5.72}$ & & & \\[2ex]  
% colour, my fit #5
    & { \boldmath $M_V \!=\! a_0 + a_1 X + a_2 Y$ } &   &   &   &   &   &   \\[1.0ex]  % title
\#6 &
$0.22 \!<\! P \!<\! 0.90$ d &       &      &     &  318   &      &    \\  
    & $X\!=\! \log{P}+0.35$         
    & $3.21$                 & $-5.70$   & $+2.49$    &     & 812.4  & 2.6 & 0.26 & 0.20 & 0.79\\ % LSQ
    & $Y\!=\!(B-V)_0\!-\!0.5$ & $3.20_{-0.05}^{+0.05}$ & $-5.77_{-0.38}^{+0.44}$     
    & $+2.54_{-0.32}^{+0.29}$    &     &     &      \\
    & $P_0=0.4467$ d \\[2ex]  % boot
% central w/colour
\#7 &
$0.275 \!<\! P \!<\! 0.575$ d &       &      &     &  297   &      &    \\  
    & $X\!=\!\log{P}+0.40$         
    & $3.82$                 & $-9.04$   & $+0.05$    &     & 1638.4  & 5.6 & 0.39 & 0.31 & 0.83\\ % LSQ
    & $Y\!=\!(B-V)_0\!-\!0.5$ & $3.81_{-0.05}^{+0.05}$ & $-8.92_{-0.81}^{+0.81}$     
    & $+0.12_{-0.45}^{+0.56}$    &     &     &    \\
    & $P_0=0.3981$ d  \\[2ex]  % boot
\enddata
\tablecomments{For each solution, the results are given twice: (1)~for a standard 
least-squares determination for $n$ stars, with the corresponding $\chi^2$, the reduced $\chi_r^2$, 
the weighted mean-square deviation $\epsilon$,  the weighted mean 
absolute deviation $\delta$ (weights were $\propto 1/\sigma_i^2$),
and the coefficient of determination $R^2$ (see the text), 
and (2)~as a result of a bootstrap experiment (10,000 repetitions) 
with the median (50th percentile) and the asymmetric errors given as
differences from the median for the 15.8th and the 84.1st 
percentiles of the coefficient distributions. 
In the column ``Details'', the orbital period $P$ is specified by its range.
Note that shifts to mid-ranges of $\log{P}$ and $(B-V)_0$ 
have been applied in the definition of the independent variables 
$X$ and $Y$ to improve the least squares solutions. 
$P_0$ is the period corresponding to the $\log{P}$ 
shift in the definition of $X$. 
}
\end{deluxetable}


\newpage

\begin{thebibliography}{}

% Acta Astronomica \!=\!\!>\! \actaa

\bibitem[Chen et al.(2016)]{Chen2016}
   Chen, X., de Grijs, R., \& Deng, L. 2016, \apj, 832, 138

\bibitem[D'Angelo et al.(2006)]{Trip2}                % triples 2
   D'Angelo, C., van Kerkwijk, M. H. \& Rucinski S. M. 
   2006, \aj, 132, 650

\bibitem[Eaton(2016)]{Eaton2016}
   Eaton, J. A. 2016, \mnras, 457, 836

\bibitem[Eggen(1967)]{Egg1967}
   Eggen, O. J. 1967, \mnras, 70, 111

\bibitem[ESA(1997)]{esa1997} 
   European Space Agency, 1997, 
   The Hipparcos and Tycho catalogs (ESA SP-1200)
   (Noor\-dwijk: ESA)
   
\bibitem[Fabricius \& Makarov(2000)]{FM2000} 
   Fabricius, C., \& Makarov, V. V. 2000, \aap, 356, 141
   
\bibitem[Gaia(2016a)]{Gaia2016a}
   Gaia Collaboration,  Brown, A. G. A., Vallenari, A., T. Prusti, T., 
   de Bruijne, J. H. J., Mignard, F., Drimmel, R., Babusiaux, C.  
   Bailer-Jones, C. A. L., Bastian, U. 
   \& et al. 2016a, \aap, 595, A2

\bibitem[Gaia(2016b)]{Gaia2016b}
   Gaia Collaboration, Prusti, T., de Bruijne, J. H. J., Brown, A. G. A., 
   Vallenari, A., Babusiaux, C., Bailer-Jones, C. A. L., Bastian, U.,
   Biermann, M., Evans, D. W., 
   \& et al. 2016b, \aap, 595, A1

\bibitem[Gettel et al.(2006)]{GGM2006}
   Gettel, S. J., Geske, M. T., \& McKay, T. A., 2006, \aj, 131, 621 (GGM2006)

\bibitem[Gray(2005)]{Gray2005}
   2005, The Observation and Analysis of Stellar Photospheres,
   Cambridge University Press (Cambridge), Appendix B. 

\bibitem[Hill et al.(1989)]{Hill1989}
   Hill, G., Fisher, W. A., \& Holmgren, D. 1989, \aap, 211, 81
   
\bibitem[H{\o}g et al.(2000)]{Hog2000}
   H{\o}g, E., Fabricius, C., Makarov, V. V., Urban, S., Corbin, T., 
   Wycoff, G., Bastian, U., Schwekendiek, P., \& Wicenec, A., 
   2000, \aap, 357, 367

\bibitem[Kaluzny(1997)]{JKa1997}                          % GC NGC6397
   Kaluzny, J. 1997, \aaps, 122, 1

\bibitem[Kaluzny et al.(1997)]{KTK1997}                   % GC M4
   Kaluzny, J., Thompson, I. B., \& Krzeminski, W. 
   1997, \aj, 113, 2219

\bibitem[Lohr et al.(2013)]{Lohr2013}
    Lohr, M. E., Norton, A. J., Kolb, U. C., Maxted, P. F. L,  Todd, I. \& West, R. G.
    2013, \aap, 549, A86

\bibitem[Lucy(1968a)]{Lucy1968a}
    Lucy, L.B. 1968a, \apj, 151, 1123

\bibitem[Lucy(1968b)]{Lucy1968b}
    Lucy, L.B. 1968b, \apj, 153, 877

\bibitem[Lutz \& Kelker(1973)]{LK1973}
    Lutz, T. E., \& Kelker, D. H. 1973, \pasp, 85, 573

\bibitem[Maciejewski et al.(2008)]{Maciej2008}
    Maciejewski, G., Boeva, S., Georgiev, Ts., Mihov, B., Ovcharov, E., 
    Valcheva, A., Niedzielski, A. 2008, BaltAstr, 17, 51

\bibitem[Marsh et al.(2017)]{Marsh2017}
    Marsh, F. M., Prince, T. A., Mahabal, A. A., Bellm, E. C., 
    Drake, A. J., \& Djorgovski, S. G.
    2017, \mnras, 465, 4678

\bibitem[Mazur et al.(1995)]{Mazur1995}                 % 45 bins in Cr261
   Mazur, B., Krzeminski, W., \& Kaluzny, J. 1995, \mnras, 273, 59 

\bibitem[Mochejska \& Kaluzny(1999)]{MK1999}
    Mochejska, B.J. \& Kaluzny, J. 1999, \actaa, 49, 351

\bibitem[Mochejska \& Kaluzny(2000)]{MK2000}
    Mochejska, B.J. \& Kaluzny, J. 2000, \actaa, 50, 105

\bibitem[Norton et al.(2011)]{Norton2011}
   Norton, A. J., Payne, S. G. , Evans, T., West, R. G., 
   et al. 
   2011, \aap, 528, A90

\bibitem[Nataf et al.(2013)]{Nat2013}
   Nataf, D. M., Gould, A., Fouque, P., Gonzalez, O. A., Johnson, J. A.,  Skowron, J.,
   Udalski, A., Szymanski, M. K., Kubiak, M., Pietrzynski, G., Soszynski, I., 
   Ulaczyk, K., Wyrzykowski, L., Poleski, R. 2013, \apj, 769, 88

\bibitem[Pawlak(2016)]{Paw2016}
   Pawlak, M. 2016, \mnras, 457, 4323
   
\bibitem[Pecaut \& Mamajek(2013)]{PM2013}
   Pecaut, M. J., Mamajek, E. E. 2013, \apjs, 208, 9

\bibitem[Pojmanski et al.(2005)]{asas}
   Pojmanski, G., Pilecki, B., \& Szczygiel, D., 2005, \actaa, 55, 275 (ASAS)

\bibitem[Pribulla \& Rucinski(2006)]{Trip1}                % triples 1
   Pribulla, T, \& Rucinski, S. M. 
   2006, \aj, 131, 2986

\bibitem[Raghavan et al.(2010)]{Rag2010} 
   Raghavan, D., McAlister, H. A., Henry, T. J., Latham, D. W., Marcy, G. W.,
   Mason, B. D., Gies, D. R., White, R. J. \& ten Brummelaar, T. A.
   2010, \apjs, 190, 1

\bibitem[Rucinski(1994)]{Rci1994}                     % calib-1
   Rucinski S. M., 1994, \pasp, 106, 462

\bibitem[Rucinski(2000)]{Rci2000}                     % glob
   Rucinski S. M., 2000, \aj, 120, 319

\bibitem[Rucinski(2002)]{Rci2002}                     % 7.5 mag
   Rucinski S. M., 2002, \pasp, 114, 1124

\bibitem[Rucinski(2006)]{Rci2006}                     % ASAS LF
   Rucinski S. M., 2006, \mnras, 368, 1319

\bibitem[Rucinski(2007)]{Rci2007}                     % ASAS short-period
   Rucinski S. M., 2007, \mnras, 382, 393

\bibitem[Rucinski(2015)]{Rci2015}                     % AW UMa
   Rucinski, S. M. 2015, \aj, 149, 49

\bibitem[Rucinski \& Duerbeck(1997)]{RD1997}          % calib
   Rucinski S. M., \& Duerbeck H. W. 1997, \pasp, 109, 1340 (RD1997)

\bibitem[Rucinski \& Paczynski(2002)]{RP2002}        % EW with trend
   Rucinski S. M., \& Paczynski, 2002, Inf.\ Bull.\ Var.\ Stars, 5321

\bibitem[Rucinski et al.(2007)]{Trip3}                % triples 3
   Rucinski S. M., Pribulla, T, \& van Kerkwijk, M. H.
   2007, \aj, 134, 2353

\bibitem[Rucinski et al.(2013)]{Rci2013}               % metallicities
    Rucinski S.M., Pribulla, T., \& Budaj, J. 2013, \aj, 146, 70

\bibitem[Schlegel et al.(1998)]{Schleg1998}
   Schlegel, D. J., Finkbeiner, D. P., \& Davis, M. 1998, \apj, 500, 525

\bibitem[Skrutskie et al.(2006)]{2mass}
   Skrutskie, M. F., et al.       % plus 30 co-authors
   2006, \aj, 131, 1163.

\bibitem[Soszy\'nski et al.(2015)] {Sosz2015}
   Soszy\'nski, I, St\c{e}pie\'{n}, K, Pilecki, B., Mr\'{o}z, P., Udalski, A., 
   et al.
   2015, \actaa, 65, 39

\bibitem[St\c{e}pie\'{n}(2006a)]{Step2006a}         % evolution I  {Step2006a,Step2006b,Step2011}
   St\c{e}pie\'{n}, K. 2006a, \actaa, 56, 199

\bibitem[St\c{e}pie\'{n}(2006b)]{Step2006b}         % evolution II
   St\c{e}pie\'{n}, K. 2006b, \actaa, 56, 347

\bibitem[St\c{e}pie\'{n}(2009)]{Step2009}          % model with global circulation
   St\c{e}pie\'{n}, K. 2009, \mnras, 397, 857

\bibitem[St\c{e}pie\'{n}(2011)]{Step2011}          % approach to contact
   St\c{e}pie\'{n}, K. 2011, \actaa, 61, 139

%\bibitem[St\c{e}pie\'{n} \& Gazeas(2011)]{SG2012}         % low mass contact
%   St\c{e}pie\'{n}, K. \& Gazeas, K. 2012, \actaa, 62, 153

%\bibitem[St\c{e}pie\'{n} \& Kiraga(2012)]{SK2012}         % contacts in GCs
%   St\c{e}pie\'{n}, K. \& Kiraga, M. 2012, \aap, 577, A117

\bibitem[Terrell et al.(2012)]{Terr2012}
   Terrell, D., Gross, J., \& Cooney, Jr., W. R.
   2012, \aj, 143, 99

\bibitem[Thompson et al.(1999)]{ITh1999}                % NGC6752
   Thompson, I. B., Kaluzny, J., Pych, W. \& Krzeminski, W.
   1999, \aj, 118, 462

\bibitem[Webbink(2003)]{Webb2003}
   Webbink, R. F. 2003, 
   3D Stellar Evolution, eds. Turcotte, S., Keller, S. C. \& Cavallo, R. M. 
   ASP Conf.\ Ser. , 293,  76

\end{thebibliography}
\end{document}